\documentstyle[12pt]{article}
 \def\de{depth}

   \let\de=\delta
   \let\th=\theta
\let\dh=\vartheta  \let\la=\lambda 
   \let\r=\rho 
\let\om=\omega 

\let\ph=\varphi \let\Ph=\phi  
\let\Om=\Omega

\def\0{\over } \def\1{\vec }     \def\2{{1\over2}} \def\4{{1\over4}}
\def\5{\bar }  \def\6{\partial } \def\7#1{{#1}\llap{/}}
\def\8#1{{\textstyle{#1}}}       \def\9#1{{\bf {#1}}}

\def\llp{\hbox to 0pt{\hss /\hskip1.5pt}}
\def\llo{\hbox to 0.2pt{\hss /}}
\def\llq{\hbox to 0pt{\hss /\hskip0.5pt}}
\def\so{\supset\hbox to 0pt{\hss $\displaystyle -$\hskip1pt}}

\def\<{\left\langle } \def\>{\right\rangle }
 
\def\[{\left\lbrack } \def\]{\right\rbrack }
\def\({\left( } \def\){\right) }

  \let\ex=\times

 \def\CM{{\cal M}}

\def\RR{ I \hspace*{-0.8ex} R }

\def\bea{\begin{eqnarray}}
\def\eea{\end{eqnarray}}
\def\beann{\begin{eqnarray*}} \def\eeann{\end{eqnarray*}}
\def\beq{\begin{equation}} \def\eeq{\end{equation}}

{\begin{list}{}%
{\settowidth{\labelwidth}{#1}%
\setlength{\leftmargin}{\labelwidth}
\addtolength{\leftmargin}{\labelsep}
}}
{\end{list}}

\def\_#1{\underline{#1}}

  \let\and=\wedge

\def\|#1{\bigg|_{_{#1}}} 
\let\th=\theta 

\def\intl#1{\int\limits#1 }
\def\ll#1{\label{#1}}

\def\strich{\mathop{\vphantom{\sum} \vert   } }
\newfont{\bit}{cmbxti10 scaled 1728}

\begin{document}
\pagestyle{empty}
\renewcommand{\thefootnote}{\fnsymbol{footnote}}
\hfill  TUW -- 93 -- 12 \\
\hfill gr-qc/9305009  \\
\vfill

\begin{center}
{\Large
On the Distributional Nature of the Energy Momentum Tensor of a Black Hole or
\\

\medskip

What Curves the Schwarzschild  Geometry ? }
\vfill

Herbert BALASIN\footnote[1]{e-mail: hbalasin @ email.tuwien.ac.at}
\\
Herbert NACHBAGAUER\footnote[7]{e-mail: nachb @ email.tuwien.ac.at}
\\
\medskip
Institut f\"ur Theoretische Physik, Technische Universit\"at Wien\\
Wiedner Hauptstra{\ss}e 8--10, A - 1040 Wien, AUSTRIA
\end{center}
\vfill

\begin{abstract} Using distributional techniques we calculate the
energy--momentum tensor of the Schwarzschild geometry. It turns out
to be a well--defined tensor--distribution concentrated on the
$r=0$ region which is usually excluded from space--time. This
provides a physical interpretation  for the curvature of this
geometry.
\end{abstract} \vfill

\hfill
\today \\
\renewcommand{\thefootnote}{\arabic{footnote}}

\pagebreak
\pagenumbering{arabic}
\pagestyle{plain}

\section*{\Large\bit 1.\ Introduction}

Recently the interest in distributional solutions of general
relativity \cite{geroch}, especially the gravitational field of
shock-waves \cite{aichelburg,thooft,lousto} was increased  by
the consideration of scattering (test) particles in these geometries.
 It turns out
that the scattering amplitudes are exactly calculable and that they
display a particular similarity  to the ones obtained in string
theory.  This analogy is made rigorous by considering the limit of
general relativity \cite{verl} where only a two--dimensional block of
the metric is quantised and the remaining part is treated
classically.

The aim of the present paper is to investigate the origin of the
shock--wave geometry which arises from boosting the Schwarzschild
metric. As already pointed out by Aichelburg and Sexl
\cite{aichelburg} the metric becomes singular at boost velocity
approaching the speed of light, even in the sense of distributions.
However, by performing a clever ``singular''
coordinate-transformation it is possible to obtain a finite result
with a sensible physical interpretation, namely an energy--momentum
tensor that is concentrated on a light-like line.  This is somewhat
puzzling since the original geometry is considered to be a vacuum
solution with zero energy--momentum tensor.  Thus one may
ask for the origin of curvature in the Schwarzschild space-time.

As we shall show the so-called vacuum solution is such that the
energy--momentum tensor is concentrated on regions usually excluded
from space-time (i.e.\ the origin in Schwarzschild coordinates),
resulting in the physically unsatisfactory situation that curvature
is generated by a zero energy--momentum tensor.

In our approach, we consider the appropriately chosen manifold as
the underlying object to be given (physical) structures such as a
metric tensor.  Thus  we include these regions of space-time in the
manifold now admitting also for distribution valued metrics.

The present work shows that this is in fact possible in the case of
the Schwarzschild geometry. To begin with we want to give a brief
sketch of the general framework of distributions on arbitrary
manifolds, since we believe that this material is
important but not quite standard \cite{lichn}.
Afterwards we treat some simple but instructive
examples. Finally we turn to the Schwarzschild geometry and
calculate its energy--momentum tensor.
\pagebreak[2]

\section*{\Large \bit 2.\ Mathematical Framework}

Usually a distribution $f$ is thought of as linear functional
acting on an appropriate test function space $C_0^\infty$ (e.g.\
$C^\infty$--functions with compact support)  \cite{gelf}.  This
definition is motivated by the need to generalise the regular
functionals that arise as integrals of locally integrable functions
with a test function $\ph$
\beq ( f,\ph ) = \int d^n x f(x) \ph (x),\quad
\ph \in C_0^\infty \ll{1}.\eeq
This definition makes implicit use of the test functions defined on
$\RR ^n$. In order to generalise the concept  to an  arbitrary
manifold $\CM $, we consider instead of test functions the space of
$C^\infty$-differential  n-forms $\tilde \ph $ with compact support
($\tilde \ph \in \Omega_0^n $) \cite{lichn}.  Regular functionals
arise now from locally integrable functions $f$ on $\CM$ in the
following way\footnote{The left--hand side of (\ref{2}) assumes the
existence of partitions of unity in order to define the integral.}:
\beq ( f , \tilde \ph ) = \intl_{\CM} f \tilde \ph \ll{2} . \eeq
This suggests to define distributions as linear functionals on the
space of n-forms $\Omega_0^n $. If $\CM$ is orientable with a
volume-form $\om$ every $\tilde \ph $ defines uniquely a function
$\ph \in C_0^\infty $ such that $\tilde \ph = \ph \om $ and
(\ref{2}) becomes
\beq ( f,\tilde \ph ) = \intl_{\CM} f \ph \, \om  ,\eeq
which coincides with (\ref{1}) by taking $\om = d^n x$.

Further generalisation is achieved by using tensor--valued n--forms
with compact support as test spaces and tensor fields to define
regular functionals
\beq ( t^I , \tilde \phi_I ) = \intl_{\CM}  t^I \tilde \phi_I  ,
\quad I \ldots  \mbox{ collection of tensor indices} .\eeq
This furnishes a manifestly coordinate--independent definition of
tensor--valued distributions (tensor--distributions) on a given
manifold. These definitions are completely independent of any
metrical structure on $\CM$.
\pagebreak

\section*{\Large\bit 3.\ Simple Examples}

To provide an illustration of the above concepts let us consider some
instructive
examples.

\noindent
The first example is given by the one--form
\beq   \om = { x \0 {x^2 + y^2} } dy -
{y \0 x^2 + y^2 } dx  \ll{2-1} ,\eeq
which is differentiable and closed on $\RR^2 \setminus \{ 0 \} $.
However, (\ref{2-1}) may also be understood as a
tensor--distribution on $\RR^2$. This is possible since both
component functions are locally integrable. I.e.\ the integrals
$$ \int {x \0 x^2 + y^2 }\, \ph ( x , y )\, d^{2}x , \quad
\int {y \0 x^2 + y^2}\, \ph ( x , y )\, d^{2}x $$
are well defined for all $\ph \in C_0^\infty ( \RR^2 ) $.  Let us
calculate the exterior derivative of $\om$ as tensor--distribution.
In order to facilitate the calculation we employ the following
regularisation
$$ \om = \lim_{\la \to 0}
\om_\la = \lim_{\la \to 0}  r^{\la -2}  \( x dy - y dx \) ,
\quad d\om_\la = \la r^{\la-2} \,  dx \wedge dy ,$$
$$ d \om =  \lim_{\la \to 0 } \( \la r^{\la -2 } \)  dx \wedge dy
= 2 \pi \de^{(2) } (x,y) \, dx \wedge dy . $$
The limit in the last equation was performed using the
decomposition of $r^{\la -2 } $
$$ ( r^{\la - 2 } , \ph ) = 2 \pi \intl_0^\infty r^{\la -1 }
S_\ph ( r ) dr = $$
$$ { 2 \pi \0 \la } \ph ( 0 ) + 2 \pi \intl_0^\infty r^{\la -2 }
\( S_\ph ( r ) - \th ( 1 - r ) \ph ( 0 ) \) r dr =: { 2 \pi \0 \la }
 \( \de , \ph \) + \( \[ r^{\la -2}  \] , \ph \) $$
with
$$ S_\ph (r ) ={1 \0 2 \pi } \intl_0^{2 \pi } d\Ph \,
\ph ( r \cos \Ph, r \sin \Ph ) $$
where the singular part has been isolated.  This shows that the
classically closed but non--exact one--form $\om$ turns out to be
a non--closed and therefore clearly non--exact one--form
distribution. The above derivation made explicit use of a
regularisation the result, however, is completely regularisation
independent, as will become clear in the slightly more complicated
next example, the $(\RR^3 , \de ) $--induced metric on a
cone.

\noindent
Using cylindrical coordinates the  explicit embedding
formula and metric are
\beq z =\th (\r )\r\cot\dh ,\quad dz = \th (\r )\cot\dh d\r ,
\quad ds^2 = d\r^2 ( 1 + \th (\r ) \cot^2\dh ) + \r^2 d\Ph^2
\ll{cone} .\eeq
Since the underlying manifold structure is that of $\RR^2 $ the
above metric may be viewed as tensor--distribution on $\RR^2$.
Classically the curvature of this metric vanishes, since one  cuts
out the region that corresponds to the tip of the cone.  However,
from the point of view of distribution theory this is not necessary
at all.  In order to calculate the distributional curvature we will
start with a regularisation of (\ref{cone}).  A geometrically
natural way of regularising the above metric is obtained by
considering the cone to be the limit of a sequence of hyperbolic
shells. These are given by the $(\RR^3 , \de ) $ embeddings $ z =
\sqrt{ \r^2 + a^2 } \cot \dh $ with the corresponding metric
$$ ds^2 = \( 1 + { \r^2 \0 \r^2  + a^2 } \cot^2 \dh \) d\r^2  +
 \r^2 d\Ph^2 . $$
Again we have $\RR^2$ as underlying manifold structure. The
calculation proceeds now in a straightforward fashion  using the
canonical dyad and spin connection
$$ e^\r = \sqrt{ 1 + { \r^2 \0 \r^2  + a^2 } \cot^2 \dh} \, d\r ,
\quad e^\Ph = \r \, d\Ph ,
\quad   \om^{\r\Ph} = - \( 1 + { \r^2 \0 \r^2  + a^2 } \cot^2 \dh
\)^{-{1 \0 2 } } d\Ph .$$
The corresponding curvature tensor and Ricci--scalar are
$$ R^{\r\Ph} = { a^2 \cos^2 \dh  \sin^2 \dh \0 ( a^2 \sin^2 \dh +
\r^2 )^2 } \,\, e^\r \wedge e^\Ph, \quad R = { 2 a^2 \cos^2 \dh
\sin^2 \dh \0 ( a^2 \sin^2 \dh + \r^2 )^2 } ,$$
where the coefficients are understood in the sense of distributions
$$ \lim_{a \to 0} \,( R , \ph ) = \lim_{a \to 0 }  2 \pi
\intl_0^\infty \r d\r S_\ph  ( \r ) { 2 a^2 \cos^2 \dh  \sin^2 \dh
\0 ( a^2 \sin^2 \dh + \r^2 )^2 } . $$
The singular part of the last integral may be split off in a
similar manner as in the previous example. This time, however, one
has to subtract the order two Taylor polynomial from  $S_\ph$.
Evaluation of the limit gives
$$ R = 2 \pi \cos^2  \dh \de ^{(2)} (x,y ) . $$
This result has the expected features: The curvature is
concentrated on the origin and vanishes if the cone degenerates to
a plane (i.e.\ if  $\dh = {\pi \0 2}$).

The above calculation made explicit use of our conception of the
cone as being the limit of a hyperboloid. It is nevertheless
possible to regularise (\ref{cone})  in a more abstract way by
replacing $\th ( \r ) $ by an arbitrary regularisation function $f
( \r ) $ which has to vanish at  $ \r =0 $.  The chosen form of
(\ref{cone}) guarantees that the prefactor of $d\r^2 $ in the
metric  equals unity for any regularisation $f ( \r ) $ at $\dh = {
\pi \0 2 }  $.  These requirements are met by $ f ( \r ) =
\r^\la,\, \mbox{Re} (  \la )> 0 $ yielding the regularised metric
$$ ds^2 =( 1 + \r^\la \cot^2 \dh ) d\r^2 + \r^2 d\Ph^2 . $$
A short calculation gives
$$ R_\la = -  {1 \0 \r } {\la \r^{\la-1} \cot^2 \dh \0
( 1 +  \r^\la \cot^2 \dh )^2 } ,$$
$$ \lim_{\la \to 0 } ( R_\la , \ph ) = - 2 \pi
\intl_0^{\infty } d\r {\la \r^{\la-1} \cot^2 \dh \0 ( 1 +  \r^\la
\cot^2 \dh )^2 } S_\ph ( \r )  = 2 \pi \cos^2 \dh \, \ph ( 0 ) ,$$
thus
$$ \lim_{\la \to 0 } R_\la = 2 \pi \cos^2 \dh \de^{(2)} ( x,y ) , $$
which coincides with the previous result.  Actually this result is
completely independent of the chosen regularising function $ f ( \r
) $ since repetition of the above steps with the metric
$$ ds^2 = ( 1 + f ( \r ) \cot^2 \dh ) + \r^2 d\Ph^2 $$
gives the same result in the pointwise limit $ f ( \r ) \to \th (
\r ) $.  Although the above examples made explicit use of a
regularisation, i.e.\ to stick to classical calculus, the results
are independent of the regularisation procedure employed.

\section*{\Large\bit 4.\ Schwarzschild--Geometry}

The usual derivation of the Schwarzschild metric takes advantage of
the fact that the metric is spherically symmetric and static. This
allows to fix the coordinates in a symmetry--appropriate manner and
leaves two arbitrary functions in the metric \cite{wald}.
Imposition of the vacuum Einstein equations determines those functions
and one obtains
\beq ds^2 = - \( 1 - { 2 m \0 r} \) dt^2 + \( 1 - { 2 m \0 r}
\)^{-1} dr^2 + r^2 d\Om^2 \ll{me1} . \eeq
As already pointed out in the introduction, the reason for the
curved geometry induced by the Schwarzschild solution (\ref{me1})
is somewhat missing.  In particular, since one explicitly used the
vacuum equations, the energy--momentum tensor of this geometry
vanishes leaving us with the question for the physical ground of
the non--Minkowskian geometry.

The situation is quite similar to  our first example, where $\om$
remained closed only when considered on the smaller manifold $\RR^2
\backslash \{0\}$ instead of $\RR^2$.  Physically, this line of
arguments may further be illustrated in the case of electrodynamics.

Let us calculate the static spherically symmetric solution of the
vacuum Maxwell equations. Stationarity implies that the vector
potential $A$ has to be time--independent, and rotational
invariance constrains $A$ to the form $A(x) = A_0(r) dt + A_r(r)
dr$ where r denotes the radial distance.  Using the residual gauge
freedom we eliminate the spatial component.  Finally, the vacuum
Maxwell equations $  d\!*\!d A = 0 $ leave us with
\beq
\Delta A_0(r) = {1 \0 r^2 } \6_r  \( r^2 \6_r A_0 (r) \) = 0 ,
\quad A_0 (r)= - { e \0 r },\ll{coul}
\eeq
where a possible integration--constant in the potential has been
dropped by imposing natural boundary conditions $ A_0 ( r \to \infty )
=0 $. $A_0(r) $ is precisely the Coulomb potential.

The solution obtained in this way has to be restricted to the
domain  $\RR \ex ( \RR^3 \backslash \{ 0 \} ) $ in order to be
differentiable in the classical sense. Let us now turn the argument
around and consider $ A ( x) $ as a distribution on $ \RR ^ 4 $.
Plugging (\ref{coul}) into the Maxwell equations now gives a
non--vanishing current density. The calculation boils down to the
evaluation of $ \Delta  (1 / r) = - 4 \pi \de^{(3)} ( x) $
producing  $ j^0 ( x)  = 4 \pi e \de^{(3) } ( x) $.  Physically
this shows that  the Coulomb solution may be regarded as the field
of a point--charge distribution located  at $ r = 0 $.

Let us now use this line of arguments to calculate  the
energy--momentum tensor of the Schwarzschild metric.
Similar to the Coulomb case
(\ref{coul}) the metric (\ref{me1}) is a well--defined
tensor--distribution even under inclusion of the line $r=0$.
Furthermore we may restrict the support of the test functions to
the ball $ r < 2 m $ which suggests to use interior
Schwarzschild--coordinates.  With regard to the experience gained
in the last two sections we choose the  regularisation
$$ ds^2 = h(r) dt^2 - {1 \0 h(r)}  dr^2 + r^2 d\Om^2 , \quad
h(r) =   -1 + { 2 m \0 r} f(r) , $$
which becomes flat space upon turning off the mass $m$.  The
calculation proceeds now in a straightforward manner, using the
canonical tetrad and spin connection
\beq e^t = \sqrt{ h(r) } dt ,\quad e^r =  h(r)^{-{1 \0 2 } } dr,
\quad e^\th = r d\dh ,\quad e^\phi = r\sin \dh d\phi, \ll{viel}\eeq
\beq { \om^t}_r = {1 \0 2 } h'(r) dt,
\quad  {\om^\dh}_r=\sqrt{h(r)} d\dh,
\quad { \om^\phi}_r = \sqrt{h(r)} \sin \dh d\phi ,
\quad {\om^\phi}_\dh = \cos \dh d\phi .  \ll{spin} \eeq
{}From  (\ref{spin}) one finds for the Ricci--tensor and the
curvature scalar, with respect to the vielbein basis (\ref{viel})
$$   R_t= {  m \0   r } f''(r) e^t ,\quad
 R_r= -{  m \0  r } f''(r) e^r ,\quad
 R_\dh =   {2 m  \0 r^2  } f'(r) e^\dh ,\quad
 R_\phi =   {2 m  \0 r^2  } f'(r) e^\phi , $$
$$ R = 2 m \( { 1 \0 r } f''(r) + {2 \0 r^2 } f'(r) \)  . $$
In order to  evaluate  these expressions we choose an explicit
regularisation function $ f( r )  = r^\la $.  Taking the
distributional limit $ \la \to 0 $ gives
$$ R(x) = 8 \pi m\de^{(3)} (x) , $$
$$ G(x) =  8 \pi T(x) = - 8 \pi m \de^{(3)} (x) \( dt \otimes \6_t
+ dr \otimes \6_r -{1 \0 2}d\dh \otimes \6_\dh  - {1 \0 2}
d\phi \otimes \6_\phi  \) . $$
The second equality employed the coordinate basis for $T(x)$ which
coincides with our choice of the tetrad (\ref{viel}). $T(x)$ and
$R(x)$ have the desired features: They are concentrated at $ r=0$
and vanish in the limit $ m \to 0 $. Physically $T(x)$ represents
the source of the gravitational field that manifests itself in the
Schwarzschild metric.

Finally, let us comment on two important issues, regularisation
independence and type of the energy--momentum tensor.

Actually the above result could have been obtained by choosing a
different regularisation.  Examples are provided by
 $ f(r)  = r^2 / ( r^2 + a^2 ) , \, a \to 0  $
or by changing the dimension of the spherical part of the
metric. (For a discussion of this approach the reader is referred
to the Appendix.)
However, all of them give the same answer. This is due to the fact
that we might have done the whole calculation keeping $f \in
C^\infty $ at will and considering the limit $ f \to 1 $ in  the
end. Let us briefly sketch this argument with regard to the
curvature scalar
$$ (R,\ph ) = 8 \pi m \intl_0^\infty \( {1 \0 r } f''(r)
+ { 2 \0 r^2 } f'(r) \) r^2 S_\ph (r), $$
where the integral can be rewritten by partial integration
$$   f'(r) r S_\ph (r) + f(r)  S_\ph (r)- f(r) r
S_\ph '(r)  \strich_0^\infty + \intl_0^\infty dr  f(r)
\[ (r  S_\ph (r) )'' - 2 f(r)  S_\ph '(r) \] . $$
One obtains, by evaluating the boundary terms, thereby taking into
account the zero of $f(r)$ at $r=0$ and the compact support of
$S_\ph (r)$ the following result
$$ 8 \pi m  \intl_0^\infty dr r S_\ph ''(r) = 8 \pi m \(
r  S_\ph '(r) \strich_0^\infty 
- \intl_0^\infty dr S_\ph '(r) \) = 8 \pi m S_\ph (0) . $$

Secondly, our calculation gave only the mixed components of the
energy--momentum tensor. Due
to the singular behaviour of the metric the other forms simply do
not exist.  However, this causes no loss of information since those
forms are redundant. Moreover, the mixed form may be considered
being the most ``democratic'' which  also allows the calculation of
the traces.

\section*{\Large\bit 5.\ Conclusion}

The present work advocates the use of distributional techniques to
calculate (geo)metrical quantities of manifolds equipped with a
singular metric. This approach is tested on simple (conceptive)
examples, like the curvature of a cone or the integrability of a
vector field on $\RR^2$. Finally, the proper subject of our work,
the energy--momentum tensor of the Schwarzschild space--time is
calculated.  One finds that it is a tensor--distribution supported
in the singular region.  From a physical point of view this allows
the identification of the source of the Schwarzschild geometry
putting the starting point of the shock--wave geometry of
Aichelburg and Sexl on the same footing with the result of the
boost.  The presented techniques may also be applied
to the Kerr geometry. Work in this direction is currently under
progress.
\vfill

\noindent{\em Acknowledgement:\/} The authors are greatly indebted
to Prof.~P.~C.~Aichelburg for many useful discussions.

\pagebreak
\section*{\Large\bf Appendix} \noindent
\section*{\Large\bit Dimensional\\ \hspace*{0.3cm} Regularisation of the
Schwarzschild Metric}
\renewcommand{\theequation}{A.\arabic{equation}}
\setcounter{equation}{0}
This regularisation is conceptually different from the ones
previously used. It takes advantage of the spherical symmetry of
the Schwarzschild space--time which may be  interpreted as
$S^2$--bundle over the two--dimensional $(r,t)$--space containing
the interesting geometry. Replacing $S^2$ by $S^{2\omega}$ the
$(r,t)$--geometry remains unchanged but the $(2\omega
+2)$--dimensional quantities such as curvature are regularised.

As usual we begin with the metric and its canonical $(2\omega +
2)$--frame
$$ ds^2 = - \( 1-\frac{2m}{r} \) dt^2 +
\( 1-\frac{2m}{r}\)^{-{1\0 2}} dr^2 + r^2 d\Omega^2_{(2\omega)},
$$
\beq e^t =\( \frac{2m}{r}-1 \)^{1 \0 2 } dt, \quad e^r =
\( \frac{2m}{r}-1 \)^{-{1 \0 2 }} , \quad e^i = r\,\tilde{e}^i,
 \label{dimsch} \eeq
where $d\Omega^2_{(2\omega)}$ and $\tilde{e}^i$ denote the
line--element of the $2\omega$--sphere and its canonical
$2\omega$--frame respectively. (In the following we will use a
tilde to denote $S^{2\omega}$--quantities like frame,
spin--connection and curvature). The spin--connection and curvature
of (\ref{dimsch}) are given by
$$ \omega^t{}_r = -\frac{m}{r^2}\,dt, \quad
\omega^i{}_r = \left(\frac{2m}{r}-1\right)^{\frac{1}{2}}\,
\tilde{e}^i, \quad
\omega^i{}_j = \tilde{\omega}^i{}_j , $$
$$   R^t{}_r = \frac{2m}{r^3}\, e^r \wedge e^t, \quad
R^t{}_i = -\frac{m}{r^2} \, e^t \wedge \tilde{e}^i, $$
$$ R^r{}_i = -\frac{m}{r^2}\,  e^r \wedge \tilde{e}^i, \quad
R^i{}_j = \tilde{R}^i{}_j +
\left( \frac{2m}{r}-1 \right) \tilde{e}^i \wedge\tilde{e}^j. $$
Taking into account that
$\tilde{R}^i{}_j=\tilde{e}^i\wedge\tilde{e}^j$ which reflects the
fact that $S^{2\omega}$ is a space of constant curvature, the
Ricci--tensor and the curvature scalar become
$$ R^t= -\frac{2m}{r^3}(\omega-1)e^t,\quad
R^r= -\frac{2m}{r^3}(\omega-1)e^r,\quad
R^i= \frac{4m}{r^3}(\omega-1)e^i, $$
$$  R_\omega = \frac{4m}{r^3}(\omega-1)(2\omega-1). $$
The distributional limits of these quantities are calculated in the
usual way.  In order to see this explicitly let us calculate the
curvature scalar.
$$ ( R,\varphi) = \lim_{\omega\to 1}(R_\omega,\varphi), $$
$$ (R_\omega,\varphi) = \int d^{2\omega+1}x R(x)\varphi(x) =
 4m (\omega-1)(2\omega-1)
  \frac{2\pi^{\omega+\frac{1}{2}}}{\Gamma(\omega+\frac{1}{2})}
 \intl_0^\infty  \frac{ r^{2\omega}}{r^3}
S_\varphi(r) dr = $$
$$ = \frac{8m(2\omega-1)\pi^{\omega+\frac{1}{2}}}
{2\Gamma(\omega+\frac{1}{2})} \left(
 \left. r^{2\omega-2}S_\varphi(r)\right|^\infty_0
- \intl_0^\infty r^{2\omega-2}
S'_\varphi(r)dr\right) , $$
where the averaged test function is defined as follows
$$ S_\varphi(r)=\frac{1}{\vert \Omega_{2\omega}\vert}
\intl_{S^{2\omega}} d\Omega\,\varphi(r\Omega),\quad
\vert\Omega_{2\omega}\vert =
\frac{2\pi^{\omega+\frac{1}{2}}}{\Gamma(\omega+\frac{1}{2})}. $$
Therefore the limit $\omega\to 1$ becomes
$$ (R,\varphi) = 8\pi m S_\varphi(0) = 8\pi m\varphi(0) $$
which coincides with our previous  result.

\end{document}